\documentclass[12pt,epsf,fleqn]{article}
\usepackage{epsfig}
\usepackage[dvips]{color}
\begin{document}
\title{Electroweak phase transition in an extension of the Standard Model with scalar color octet}
\author{S. W. Ham$^{(1,2)}$\footnote{s.w.ham@hotmail.com},
Seong-A Shim$^{(3)}$\footnote{shims@sungshin.ac.kr},
and S. K. Oh$^{(4)}$\footnote{sunkun@konkuk.ac.kr}
\\
\\
{\it (1) School of Physics, KIAS, Seoul 130-722, Korea} \\
{\it (2) Korea Institute of Science and Technology Information} \\
{\it Daejeon, 305-806, Korea} \\
{\it (3) Department of Mathematics, Sungshin Women's University} \\
{\it Seoul 136-742, Korea} \\
{\it (4) Department of Physics, Konkuk University, Seoul 143-701,
Korea}
\\
\\
}
\date{}
\maketitle
\begin{abstract}
In an extension of the Standard Model with a scalar color octet,
the possibility of the strongly first-order electroweak phase transition is studied,
by examining the finite-temperature effective Higgs potential at the one-loop level.
It is found that there are wide regions in the parameter space that allow
the strongly first-order electroweak phase transition, where
the Higgs boson mass is larger than the experimental lower bound of 115 GeV,
and the masses of the scalar color octet is around 200 GeV.
The parameter regions may be explored at the LHC with respect to the electroweak phase transition.
\end{abstract}
\vfil\eject

\section{Introduction}

The observed baryon asymmetry of the universe, or the excess of matter over anti matter,
is a challenging problem for any theoretical model to be phenomenologically realistic.
Several decades ago, Sakharov pointed out that theoretical models can generate the baryon asymmetry
dynamically if they satisfy three essential conditions:
the violation of baryon number conservation, the violation of both C and CP,
and the deviation from thermal equilibrium [1].
It is well known that, in order to ensure sufficient deviation from thermal equilibrium, the electroweak
phase transition (EWPT) should be first order, and its strength should be strong,
since otherwise the baryon asymmetry generated during the phase transition subsequently would disappear [2-10].
As the universe cools down, the shape of the potential of the scalar field that is responsible for the electroweak
symmetry breaking has two degenerate minima, where one of them is the (false) vacuum of the symmetric state
and the other is the (true) vacuum of the broken state, at the critical temperature.
The first-order electroweak phase transition takes place from the false vacuum to the true vacuum.
In general, the first-order electroweak phase transition
is regarded as strong if the vacuum expectation value (VEV)
of the scalar field at the true vacuum is larger than the critical temperature.

The Standard Model (SM) is certainly the most successful theory so far for electroweak interactions,
yet it is found, however, that the SM faces severe difficulty to satisfy the Sakharov conditions.
First, the complex phase in the Cabibbo-Kobayashi-Maskawa (CKM) matrix cannot produce large enough
CP violation to generate the baryon asymmetry.
Further, for the present experimental lower bound on the mass of the SM Higgs boson,
the strength of the first-order EWPT in the SM is too weak.
Consequently, the possibility of a strongly first-order EWPT may be studied in the
models modified or extended the SM in order to explain the baryon asymmetry of the universe.

In the literature, a number of models alternative to the SM have been investigated in this context.
These models are phenomenologically well motivated, if not necessarily motivated by
the baryon asymmetry.
We are interested in studying the possibility of a strongly first-order EWPT in these models.
Among them is an extension of the SM with an additional scalar color octet.
Popov, Povarov, and Smirnov have studied it within the context of Pati-Salam unification [11].
Manohar and Wise have studied the general structure of this model,
and examined new impact on the Higgs phenomenology and flavor physics [12].
Recently, other authors have also investigated the implications of the scalar color octet
at the CERN Large Hadron Collider (LHC) [13-19].

It is also noticed that the presence of the scalar color octet causes additional sources of CP violation
beyond the CKM matrix in the SM [12].
Since a sufficient CP violation is required by one of the Sakharov conditions for generating
the baryon asymmetry, this model is in a better position than the SM in this respect.
We are thus interested in whether this model also allows a strongly first-order EWPT.

By studying the finite temperature effective Higgs potential at the one-loop level,
we find that there are parameter regions in this model where the EWPT is
strongly first order to generate the desired baryon asymmetry.
In the parameter regions, the Higgs boson mass can be as large as 197 GeV
and the masses of the scalar color octet are as large as about 250 GeV.
Thus, these parameter regions may be explored at the LHC.

\section{The Model}

In Ref. [12], Manohar and Wise have considered the generalization of the SM with
the most general scalar sector, with a natural suppression of flavor changing neutral currents.
Explicitly, they have considered the case of an additional scalar color octet, from the point of
view of the LHC phenomenology.
Let us briefly describe the model.

The scalar sector of this model consists of $H$, the usual SM Higgs doublet,
and $S^\alpha$ ($\alpha = 1, \cdot, 8$), the scalar color octet.
The SM Higgs doublet is defined as $H^T = (H^+, H^0)$,
where $H^+$ and $H^0$ are the charged and neutral Higgs fields, respectively.
The scalar color octet are defined as
\begin{eqnarray}
S^\alpha= \bigg(
\begin{array}{c}
S^{+\alpha} \\
S^{0\alpha}
\end{array} \bigg)
= \bigg (
\begin{array}{c}
S^{\alpha}_C \\
{\displaystyle S_R^{\alpha} + i S_I^{\alpha}  \over \displaystyle \sqrt{2} }
\end{array}
\bigg )   \ ,
\end{eqnarray}
where $S_C^\alpha$ is the charged scalar color octet and
$S_R^{\alpha}$ and $S_I^{\alpha}$ are respectively the real and the complex components of the
neutral scalar color octet.

Under $SU(3) \times SU(2) \times U(1)$, the SM Higgs field transforms as
$({\bf 1}, {\bf 2})_{1/2}$ and the scalar color octet as $({\bf 8}, {\bf 2})_{1/2}$.
The Yukawa coupling of the SM Higgs boson to the SM quarks is given as
\begin{equation}
{\cal L} = - g_{ij}^U {\bar u}_{R i} Q_{L j} H
- g_{ij}^D {\bar d}_{R i} Q_{L j} H^{\dagger} + {\rm H.c.} \ ,
\end{equation}
where $i$ and $j$ are flavor indices ($i,j = 1,2,3$),
$g_{ij}^U$ and $g_{ij}^D$ are the Yukawa coupling coefficients,
$Q_{Lj}$ are the quark doublets,
and $u_{Ri}$ and $d_{Ri}$ are the quark singlets.
The Yukawa couplings of the scalar color octet to the SM quarks are given as
\begin{equation}
{\cal L} = - \eta_U g_{ij}^U {\bar u}_{R i} T^\alpha Q_{L j} S^\alpha
- \eta_D g_{ij}^D {\bar d}_{R i} T^\alpha Q_{L j} S^{\alpha \dagger} + {\rm H.c.} \ ,
\end{equation}
where $\eta_U$ and $\eta_D$ are generally complex constants,
$\alpha$ is the color index ($\alpha = 1, \cdot, 8$),
$T^\alpha$ are the $SU(3)$ generators with the normalization condition of
${\rm Tr} (T^\alpha T^\beta) = \delta^{\alpha\beta}/2$.

In terms of the usual SM Higgs doublet and the scalar color octet,
the most general form of the scalar potential at the tree level at zero temperature is given as [12]
\begin{eqnarray}
V_0 &=& -\mu^2 H^{\dagger i} H_i + \lambda \left(H^{\dagger i} H_i \right)^2  \cr
&& + 2 m_S^2 {\rm Tr} S^{\dagger i} S_i
+ \lambda_1 H^{\dagger i} H_i {\rm Tr} S^{\dagger j} S_j
+ \lambda_2 H^{\dagger i} H_j {\rm Tr} S^{\dagger j} S_i \cr
&& + \Bigl[ \lambda_3  H^{\dagger i} H^{\dagger j} {\rm Tr} S_i S_j
+ \lambda_4 H^{\dagger i} {\rm Tr} S^{\dagger j} S_j  S_i
 + \lambda_5 H^{\dagger i} {\rm Tr} S^{\dagger j} S_i  S_j + {\rm H.c.} \Bigr]  \cr
&&+ \lambda_6 {\rm Tr} S^{\dagger i} S_i S^{\dagger j} S_j
+ \lambda_7 {\rm Tr} S^{\dagger i} S_j S^{\dagger j} S_i
+ \lambda_8 {\rm Tr} S^{\dagger i} S_i  {\rm Tr} S^{\dagger j} S_j  \cr
&& + \lambda_9 {\rm Tr} S^{\dagger i} S_j  {\rm Tr} S^{\dagger j} S_i
+\lambda_{10} {\rm Tr}  S_i   S_j  {\rm Tr} S^{\dagger i} S^{\dagger j}
+ \lambda_{11} {\rm Tr}  S_i  S_j  S^{\dagger j} S^{\dagger i} \ ,
\end{eqnarray}
where $i, j$ are the $SU(2)$ indices,
traces are over the color $SU(3)$ indices.
$S = S^\alpha T^\alpha$,
$m^2_S$ is the mass parameter for the scalar color octet,
$\lambda$ is the quartic coupling coefficient of the SM Higgs field,
and ${\lambda}_i$ ($i=$ from 1 to 11) are the quartic coupling coefficients of the scalar color octet.
Note that the phases of $\eta_U$, $\eta_D$, $\lambda_4$ or $\lambda_5$ might be
additional sources of CP violation beyond the complex phase in the CKM matrix.

In this model, since the color symmetry is intact, the scalar color octet would not develop any VEV.
Thus, the electroweak symmetry breaking is triggered by the Higgs doublet alone,
and the EWPT is determined by the shape of the Higgs potential.
We assume that the real component of the neutral Higgs field, ${\rm Re}H^0 $,
develops the VEV.
Let us introduce the physical Higgs boson $\phi$ as $\phi = {\rm Re}H^0 /\sqrt{2}$.
The tree-level zero-temperature Higgs potential may then be written as
\begin{equation}
    V_0 (\phi, 0) = -{1\over 2} \mu^2 \phi^2 + {1\over 4} \lambda \phi^4   \   .
\end{equation}
At the tree level at zero temperature, we would have $\langle \phi \rangle = v_0$,
where $v_0 = 246$ GeV, the tree-level VEV.
The tree-level masses of the gauge bosons $W$, $Z$, and top quark $t$, respectively, are given as
$m_W = g_2 v_0 /2$, $m_Z = \sqrt{g_1^2 + g_2^2} v_0 /2$, and $m_t = h_t v_0 /\sqrt{2}$.
The tree-level masses of the Higgs boson and Goldstone boson are given by
$m_{\phi} = \sqrt{2 \lambda} v_0$ and $m_G = \sqrt{\lambda} v_0$, respectively.
Also, the tree-level masses for $S^{\alpha}_C$, $S_R^{\alpha}$, and $S_I^{\alpha}$
are given respectively as
\begin{eqnarray}
& & m^2_{S_C} = m_S^2+\lambda_1{ v^2_0 \over 4} \  ,  \cr
& & m^2_{S_R} = m_S^2+\left(\lambda_1+\lambda_2+2\lambda_3\right){v^2_0 \over 4}  \  , \cr
& & m^2_{S_I} = m_S^2+\left(\lambda_1+\lambda_2-2\lambda_3\right){v^2_0 \over 4}  \ .
\end{eqnarray}

At the one-loop level, the zero-temperature Higgs potential is given by including the
one-loop contributions, which is calculated, by using the effective potential method [20], as
\begin{equation}
 \Delta V_1(\phi, 0) = 2 B v^2_0 \phi^2 - {3 \over 2} B \phi^4 +
    B \log \bigg ( { \phi^2 \over v^2_0 } \bigg ) \phi^4  \  ,
\end{equation}
where
\begin{equation}
B = {3 \over 64 \pi^2 v^4_0} \bigg ( 2 m_W^4 + m_Z^4 - 4 m_t^4
+ {16 \over 3} m_{S_C}^4 + {8 \over 3} m_{S_R}^4 + {8 \over 3} m_{S_I}^4
+ {1 \over 3} m_{\phi}^4 + m_G^4 \bigg )  \  .
\end{equation}
In the radiative corrections, we include the loops of $W$ boson, $Z$ boson, top quark, the Higgs boson,
the Goldstone boson,
the charged scalar color octet, and the neutral scalar color octet.
The full one-loop zero-temperature Higgs potential is therefore given as
\begin{equation}
    V_1(\phi, 0) = V_0 (\phi, 0) + \Delta V_1 (\phi, 0)   \   .
\end{equation}

Now, the finite-temperature contribution at the one-loop level to the Higgs potential is given as [21]
\begin{eqnarray}
V_1 (\phi, T) & = &
    \sum_{l = B, F} {n_l T^4 \over 2 \pi^2}
    \int_0^{\infty} dx \ x^2 \ \log
    \bigg [1 \pm \exp{ \bigg ( - \sqrt {x^2 + {m_l^2 (\phi)/T^2 }} \bigg )  } \bigg ] ,
\end{eqnarray}
where the negative sign is for bosons ($B$) and the positive sign for fermions ($F$).
$m_l (\phi)$ is the field-dependent tree-level mass of the participating $l$-th particle,
and $n_W = 6$, $n_Z = 3$, $n_t = - 12$, $n_{\phi} = 1$, $n_G = 3$,
and $n_{S_C} = 2 n_{S_R} = 2 n_{S_I} = 16$
for the degrees of freedom for each particle, including the color factor of 8.
The full one-loop finite-temperature Higgs potential is therefore given as
\begin{equation}
    V (\phi, T) = V_1 (\phi, 0) + V_1 (\phi, T)  \ .
\end{equation}
We note that the one-loop corrected VEV of the Higgs field at zero temperature is given by
the minimum condition
\[
    {d V_1(\phi, 0) \over d \phi} = 0  \ ,
\]
and the one-loop corrected mass of the Higgs boson is given by
\[
    m^2_H = \left. {d^2 V_1(\phi, 0) \over d \phi^2} \right |_{\phi = v}  \ .
\]

For qualitative discussions on the EWPT, we take the high temperature approximation of $V_1(\phi, T)$.
It is known that in the SM the high temperature approximation is consistent
with the exact numerical integration within 5 \% at temperature $T$ for $m_F/T < 1.6$
and $m_B/T < 2.2$, where $m_F$ and $m_B$ are the mass of the relevant fermion and boson, respectively.
We assume that a similar level of accuracy may be expected in our case.
Explicitly, $V_1(\phi, T)$ is given in the high temperature approximation as
\begin{equation}
     V_1^H (\phi, T) \simeq (D T^2 - E) \phi^2 - F T \phi^3 + G \phi^4  ,
\end{equation}
where
\begin{eqnarray}
&& D = {1 \over 24 v^2} (\sum_B n_B m_B^2 + 6 m_t^2 ) \ , \cr
&& E = {m_H^2 \over 4} - {1 \over 32 \pi^2 v^2} ( \sum_{l=B,F} n_l m_l^4) \ , \cr
&& F = {1 \over 12 \pi v^3} (\sum_B n_B m_B^3) \ , \cr
&& G = {m_H^2 \over 8 v^2}
    - {1 \over 64 \pi^2 v^4} \left [\sum_{l=B,F} n_l \log {m_l^2 \over a_l T^2} \right ] \ ,
\end{eqnarray}
with $\log (a_F) = 1.14$, $\log (a_B) = 3.91$,
and $n_W = 6$, $n_Z = 3$, $n_t = - 12$, $n_{\phi} = 1$, $n_G = 3$,
and $n_{S_C} = 2 n_{S_R} = 2 n_{S_I} = 16$
for the degrees of freedom for each particle, where the color factor of 8 is taken into account.

From this formula for $V_1(\phi, T)$ in the high-temperature approximation,
one may notice that $V_1(\phi, T) \simeq D T^2 \phi^2$ at very high temperature,
and that $V_1(\phi, T) \simeq -E\phi^2 +G \phi^4$ at very low temperature.
Therefore, the term proportional to $F$ is crucial at intermediate temperature for the EWPT.
We note that contributions from the scalar color octet loops are present in $V_1(\phi, T)$.
In particular, the strength of the first-order EWPT is enhanced by the term proportional to $F$ due to
the scalar color octet contributions.
If the contributions of the scalar color octet are neglected,
$V_1 (\phi, T)$ would contain the contributions of the weak gauge bosons and top quark alone,
thus would become exactly equivalent to the SM Higgs potential.
In this case, the EWPT in this model would also become either weakly first order or higher order.
Consequently, the contributions from the loops of the scalar color octet are important for this model
to realize the strongly first-order EWPT.
For the numerical analysis in the following section, however, we perform the exact numerical integration
of $V_1(\phi, T)$.

\section{Numerical Analysis}

Let us first examine if the relevant parameters of the Higgs potential diverge at high energy scale.
In order to do so, we consider the renormaliztion group (RG) equations for them.
We know that $V_1(\phi, 0)$ contains three parameters $\lambda_1$, $\lambda_2$, and
$\lambda_3$, through the contributions of the scalar color octet, as well as the SM parameters:
$g_1$, $g_2$, and $g_3$, which are the $U(1)$, $SU(2)$, and $SU(3)$ gauge coupling constants,
respectively, and $h_t$, which is the Yukawa coupling coefficient of top quark.
These parameters are generally renormalizable, and thus should satisfy the RG equations.

Explicitly, the RG equation for the quartic coupling coefficient of the SM Higgs field, $\lambda$, is given as [15]
\begin{equation}
16\pi^2 \frac{d \lambda}{dt} = 24 \lambda^2 + 48 \lambda_1^2 + 16 \lambda_2^2 + 16\lambda_3^2
-(3 g_1^2+ 9 g_2^2 - 12 h_t^2)\lambda + \frac{3}{8}g_1^4 + \frac{3}{4} g_1^2 g_2^2
+ \frac{9}{8} g_2^4 - 6 h_t^4   \ ,
\end{equation}
where $t = \log{\mu}$, with $\mu$ being the running mass,
and the RG equations for the quartic coupling coefficients of the scalar color octet, $\lambda_1$,
$\lambda_2$, and $\lambda_3$, are given as [15]
\begin{eqnarray}
16\pi^2 \frac{d\lambda_1}{dt} &=& 16 \lambda_1^2 + 8 \lambda \lambda_1
- (\frac{3}{2}g_1^2+ \frac{9}{2} g_2^{2} - 6 h_t^2)\lambda_1
- (\frac{3}{2}g_1^2 + \frac{9}{2} g_2^2 + 18 g_3^{2}) \lambda_1  \cr
&&\mbox{} + \frac{3}{8} g_1^4 + \frac{3}{8} g_2^4
+ \frac{3}{4} g_1^2 g_2^2   \ ,   \cr
16\pi^2 \frac{d \lambda_2}{dt} & = & 8 \lambda_2^2 + 16 \lambda \lambda_2
- (\frac{3}{2}g_1^2 + \frac{9}{2} g_2^2 - 6 h_t^2) \lambda_2
- (\frac{3}{2} g_1^2 + \frac{9}{2} g_2^2 + 18 g_3^2) \lambda_2     \  ,   \cr
16\pi^2 \frac{d \lambda_3}{dt} & = & 8 \lambda_3^2 + 16 \lambda \lambda_3
- (\frac{3}{2}g_1^2 + \frac{9}{2} g_2^{2} - 6 h_t^2)\lambda_3
- (\frac{3}{2}g_1^2 + \frac{9}{2} g_2^2 + 18 g_3^{2}) \lambda_3   \ ,
\end{eqnarray}
where terms proportional to $\lambda_i$ ($i=4, \cdots,11$) are neglected,
and the contributions of $\eta_U$, $\eta_D$ and the bottom quark sector are ignored.
There are also RG equations for $g_1$, $g_2$, $g_3$, and $h_t$, which are not shown.
These RG equations exhibit the mixing terms between the SM Higgs fields and the scalar color octet.
These mixing effects may diverge at very low energies.

We calculate numerically the above RG equations, using the Runge-Kutta method.
It is found that the three SM gauge coupling coefficients, $g_1$, $g_2$, and $g_3$,
do not have any Landau poles for the whole region of the running mass up to the Planck scale.
Also, the quartic coupling coefficient $\lambda$ and the Yukawa coupling coefficient $h_t$ are seen to
increase much more slowly than the quartic coefficients of the scalar color octet, $\lambda_i$ ($i = 1,2,3$),
as the running mass increases.
Thus, we concentrate on the existence of the Landau poles of the quartic coupling coefficients
of the scalar color octet.

For the sake of simplicity, we set hereafter $\lambda_3 = 0$, thus $m_{S_R} = m_{S_I}$, and
neglect $\lambda_i$ ($i=4, \cdots,11$).
Thus, we are left with $\lambda_1$ and $\lambda_2$,
that account for the scalar color octet contributions.

We assign an initial value for both $\lambda_1$ and $\lambda_2$ at the electroweak scale,
and then let them evolve from the electroweak scale $10^2$ GeV to the unification scale over $10^{12}$ GeV,
through their respective RG equations.
If any one of them diverges in between the two scales, we set the Landau poles for both of them.
In this way, we examine the Landau poles for $\lambda_1$ and $\lambda_2$, from 0 to $4\pi$,
where $4\pi$ is set by the perturbative boundary value of the quartic coupling coefficients.
We assume that the masses of the scalar color octet are larger than 200 GeV.
We take $h_t(m_Z) = 1$ and the SM gauge coupling coefficients at $Z$ boson mass scale.

Our result for $m_H = 120$ GeV is shown in Fig. 1,
where a curve of the Landau poles is established.
For given initial value of $\lambda_1 = \lambda_2$ at $\mu =10^2$ GeV,
the curve shows the value of $\mu$ beyond which $\lambda_1$ or $\lambda_2$ becomes
divergent.
Thus, the area of Fig. 1 is divided by a boundary of the Landau poles.
The lower region of Fig. 1 is free of divergence, whereas the upper region is nonpertabative.

One may notice that if we start with a larger initial value for $\lambda_1$ and $\lambda_2$, the Landau
pole occurs at a smaller running mass
If the initial value for $\lambda_1$ and $\lambda_2$ is smaller than 0.4, the RG equations show
no divergence for the whole range of the running mass from $10^2$ GeV to $10^{12}$ GeV.
We note that, for the initial value of $\lambda_1 = \lambda_2 = 1$, the Landau pole occurs when the
running mass is a few TeV.

Another result for $m_H = 200$ GeV is shown in Fig. 2, where the values of other parameters are the same
as Fig. 1.
In Fig. 2, for the same initial value of $\lambda_1 = \lambda_2$ at the electroweak scale,
it may be observed that the Landau pole occurs at a comparatively smaller running mass than Fig. 1,
if the initial value is taken between about 0.1 and 1.

Now, we study the EWPT in this model.
For a given set of parameter values, we examine the shape of $V(\phi,T)$, by varying $T$.
If the Higgs potential exhibits the typical shape for the first-order EWPT,
with two degenerate minima and a potential barrier between them, at a certain temperature,
we define the temperature as $T_c$, the critical temperature.
We then calculate the distance between the two degenerate minima, which is defined as $v_c$,
the critical VEV, and determine the strength of the first-order EWPT. $v_c/T_c$.
In this way, we examine the parameter space of this model for the possibility of the strongly first-order EWPT.

A result is shown in Fig. 3, where $m_H = 120$, $\lambda_1 = \lambda_2 = 1$, and $m_S = 160$ GeV.
These parameter values yield $m_{S_C} = 201$ GeV and  $m_{S_R} = m_{S_I} = 236$ GeV.
We find that the Higgs potential has the shape for the first-order EWPT at the temperature $T_c = 99.7$ GeV.
The corresponding critical VEV is calculated to be $v_c = 210$ GeV.
Hence, the strongly first-order EWPT, since $v_c/T_c = 2.1$.

In Fig. 4, we show another result for a different set of parameter values,
$m_H = 120$, $\lambda_1 = \lambda_2 = 1$, and $m_S = 200$ GeV,
where the value of $m_S$ is changed.
The masses of the scalar color octet are obtained as
$m_{S_C} = 235$ GeV and  $m_{S_R} = m_{S_I} = 265$ GeV.
The critical temperature for these parameter values is $T_c = 62.1$ GeV and
the corresponding critical VEV is $v_c = 241$ GeV.
Thus, for these parameter values, too, the EWPT is strongly first-order, since $v_c/T_c = 3.8$.
The difference between Fig. 3 and Fig. 4 may be attributed to the change in $m_S$.

We now change the parameter value of $m_H$.
For $\lambda_1 = \lambda_2 = 1$, and $m_S = 160$ GeV,
we determine the critical temperature and the corresponding critical VEV for $m_H > 115$ GeV,
and calculate the strength of the first-order EWPT.
We find that for $m_H$ up to 163 GeV, the first-order EWPT is strong enough,
in other words, $v_c/T_c > 1$.
The result is shown in Fig. 5, where $v_c/T_c$ is plotted as a function of $m_H$, as a solid curve.

Also, for $\lambda_1 = \lambda_2 = 1$, and $m_S = 200$ GeV,
we do the same calculation by changing $m_H$.
We find that the EWPT may be strongly first-order, for $115 < m_H < 193$ GeV.
The result is shown in the same Fig. 5, as a dashed curve.
Therefore, Fig. 5 tells that the EWPT in this model may be strongly first-order for $115 < m_H < 163$ GeV,
$\lambda_1 = \lambda_2 = 1$, and $m_S = 160$ GeV, as well as
for $ 115 < m_H < 193$ GeV, $\lambda_1 = \lambda_2 = 1$, and $m_S = 200$ GeV.
Note that the lower bound on $m_H$ is set by the present Higgs search result, not by numerical analysis.

We also examine other regions in the parameter space of this model.
Let us set $\lambda_1 = \lambda_2 = 0.05$.
At this small value, $\lambda_1$ and $\lambda_2$ are free of the Landau poles up to
$\mu = 10^{12}$ GeV, as the results of the RG equations show.
With this value, we repeat the numerical analysis.
The results are shown in Figs. 6, 7, and 8.
Let us briefly describe them.

In Fig. 6, the shape of the Higgs potential at $T_c = 113.5$ GeV is shown,
for $m_H = 120$ GeV, and $m_S = 200$ GeV.
The critical VEV is obtained as $v_c = 183$ GeV, and the strength of the first-order EWPT is 1.6.
The masses of the scalar color octet are calculated as
$m_{S_C} = 201$ GeV and $m_{S_R} = m_{S_I} = 203$ GeV.

In Fig. 7, $m_S$ is changed to 250 GeV, while other parameter values are fixed.
The critical temperature is $T_c = 60$ GeV,
the critical VEV is $v_c = 246$ GeV,
the strength of the first-order EWPT is 4.0,
and the scalar color octet masses are $m_{S_C} = 251$ GeV and $m_{S_R} = m_{S_I} = 253$ GeV,
for $m_H = 120$ GeV, and $m_S = 250$ GeV.

In Fig. 8, the strength of the first-order EWPT is plotted as a function of $m_H$.
The solid curve is obtained for $m_S = 200$ GeV, and the dashed curve for $m_S = 250$ GeV.
These curves show that this model allows the strongly first-order EWPT for
$115 < m_H < 146$ GeV, $\lambda_1 = \lambda_2 = 0.05$, and $m_S = 200$ GeV, as well as
for $ 115 < m_H < 197$ GeV, $\lambda_1 = \lambda_2 = 0.05$, and $m_S = 250$ GeV.

Since the sizable quartic couplings to the Higgs,
$\lambda_1$ and $\lambda_2$, are crucial for allowing for a strong first-order phase transition,
we plot in Fig. 9 the strength of the first-order EWPT versus $\lambda_1/\lambda_2$ for
some values of $\lambda_2$: $\lambda_2 = 0.05$,
$\lambda_2 = 0.1$, and $\lambda_2 = 0.5$.
The values of the other free parameters are the same as in Fig. 6.
As an illustration, we obtain that the strength of the phase transition is $v_c/T_c=1.63$ for
$\lambda_1=0.05$ and $\lambda_2=0.1$
whereas $v_c/T_c=1.66$ for $\lambda_1=0.1$ and $\lambda_2=0.05$.

One may notice in Fig. 9 that the strength of the phase transition increases
as $\lambda_1/\lambda_2$ increases.
For given $\lambda_2$, $v_c/T_c$ increases as $\lambda_1$ increases.
Also, for given $\lambda_1$, $v_c/T_c$ increases as $\lambda_2$ increases.
However, comparing the three curves in Fig. 9, one may induce that the increasing rate of $v_c/T_c$
depends much strongly on $\lambda_1$ than $\lambda_2$.
This is mainly due to the fact that, as one may see in Eq. 6, $m_{S_C}$ does not depend on $\lambda_2$.

Recently, there are some extensions of the SM, in which discussions on the electroweak phase transitions
are presented [22-24].
The model with a number of additional Higgs siglets, Ref.[24], may have similar effects on the EWPT due to
the couplings between Higgs siglets to the SM Higgs doublet.
However, we note that we have the charged scalar color octet in the present model
as well as neutral Higgs scalar boson,
whereas there is no scalar color octet in the models with additional Higgs singlets.
Thus, the search for the charged scalar color octet would be helpful to distinguish the present model
from the model with additional Higgs singlets.
Also, by examining the Higgs productions via the the scalar color octet loop through the
gluon fusion process at the LHC may provide the distinctions among various models.

\section{Conclusions}

The extension of the SM with scalar color octet is significantly different from the SM
with respect to the EWPT.
In order to activate the strongly first-order EWPT, the SM requires
a very light Higgs boson, well below the experimental lower bound 114.4 GeV.
In other words, the strongly first-order EWPT is practically not allowed in the SM.
The existence of the scalar color octet in the SM improves the situation considerably,
since the strongly first-order EWPT is possible for $m_H > 115$ GeV, as our numerical analysis shows.
The thermal loop contributions to the Higgs potential at the one-loop level, given by the scalar color octet,
may play quite remarkable role on the strength of the EWPT.
In other words, the first-order EWPT might become stronger due to the thermal contributions by the scalar
color octet.

Our numerical analysis suggests that there are wide regions in the parameter space of this model
where the strongly first-order EWPT is allowed.
The allowed parameter regions are established where the mass of the Higgs boson may be consistent with
the present experimental lower bound ($m_H >115$ GeV), the masses of the scalar color octet are
within the reach of the forthcoming LHC ($m_S \simeq 200$ GeV),
and the quartic coupling coefficients for the scalar color octet are free of Landau poles.

We would like to note that we simplify our calculations by neglecting
the quartic coupling coefficients $\lambda_i$ ($i = 3, \cdots, 11$), and
$\eta_U$ and $\eta_D$ in the RG equations.
It is known that the presence of these parameters would impose a stricter bound on the running mass
coming from the perturbative-theoretic considerations such that
the Landau poles would appear at a lower running mass.
We note these simplifications are consistent with the parameter region we consider.

On the other hand, we find that the contributions of Higgs boson loops and the Goldstone boson loops
at the one-loop level are negligibly small in the present parameter region.
This is mainly because the contributions due to the Higgs boson loops are smaller than the contributions due to
the scalar color octet, in particular when the Higgs boson mass is smaller than 200 GeV and the masses
of the scalar color octet are larger than 200 GeV.

Summarizing, we establish the possibility of a strongly first-order EWPT, for the electroweak baryogenesis,
in the extension of the SM with scalar color octet.

\section*{Acknowledgments}

S. W. Ham thanks S. Baek, P. Ko, and Chul Kim for valuable comments.
He would like to acknowledge the support from KISTI under
"The Strategic Supercomputing Support Program (No. KSC-2008-S01-0011)"
with Dr. Kihyeon Cho as the technical supporter.
This research was supported by Basic Science Research Program
through the National Research Foundation of Korea (NRF) funded
by the Ministry of Education, Science and Technology (2009-0086961).



\vfil\eject

\renewcommand\thefigure{1}
\begin{figure}[t]
\begin{center}
\includegraphics[scale=0.6]{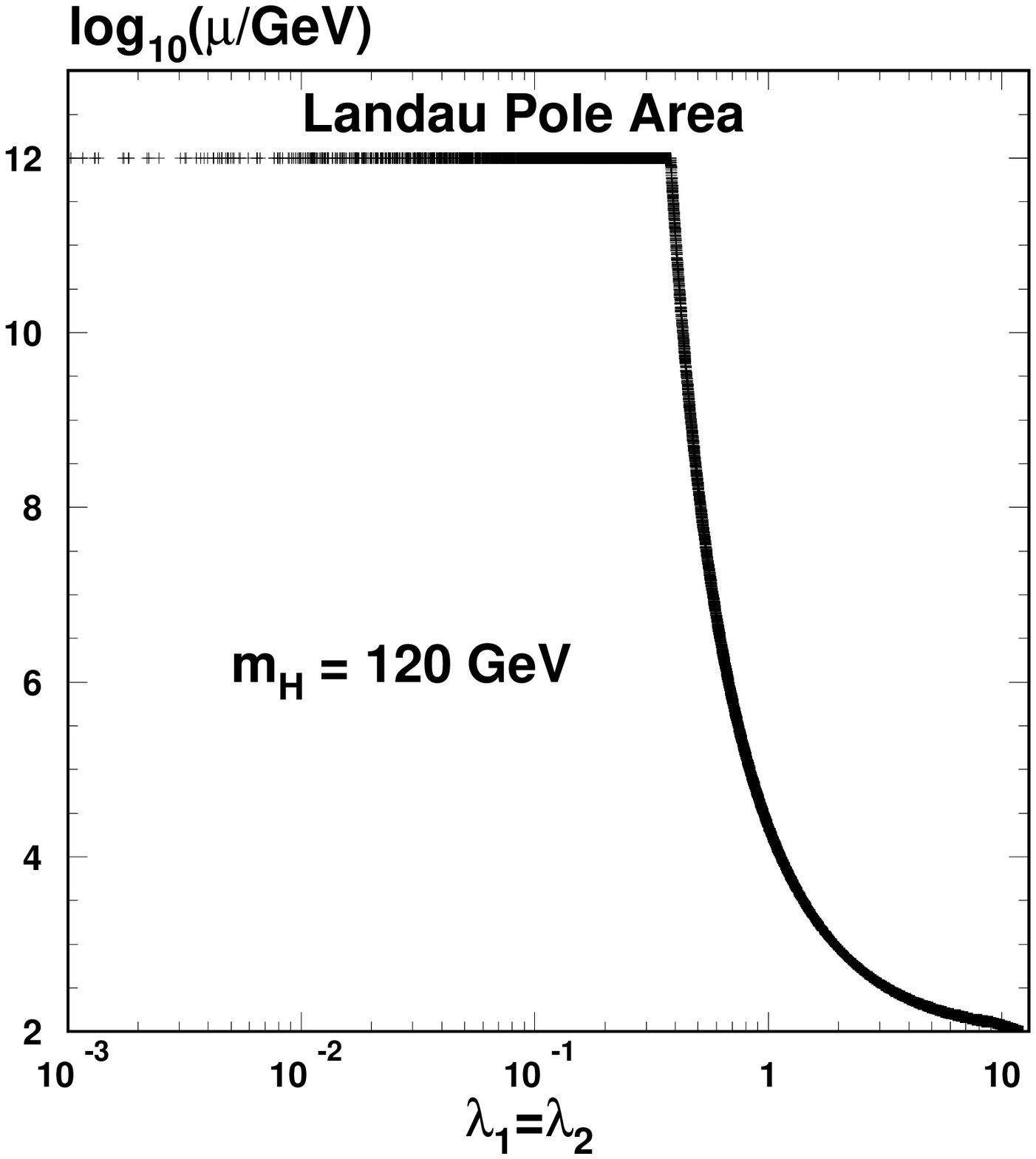}
\caption[plot]{The perturbative boundary of $\mu$, for $m_H = 120$ GeV.
For given $\lambda_1 = \lambda_2$, it is nonperturbative if $\mu$ is larger than the boundary.
We employ the RG equations for $\lambda_1$ and $\lambda_2$ to determine the value of
$\mu$ where one of them encounters the Landau pole,
with initial values of $\lambda_1 = \lambda_2$ between 0 and $4\pi$, at the electroweak scale,
together with the three gauge coupling coefficients and $h_t(m_Z) =1$.
}
\end{center}
\end{figure}

\renewcommand\thefigure{2}
\begin{figure}[t]
\begin{center}
\includegraphics[scale=0.6]{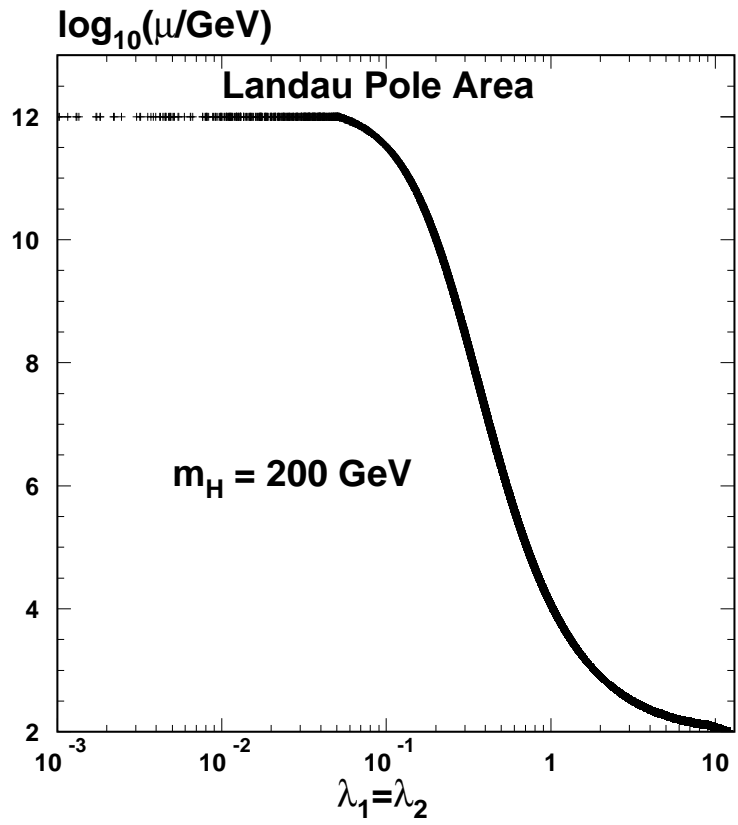}
\caption[plot]{The same as Fig. 1 except for $m_H = 200$ GeV}
\end{center}
\end{figure}

\renewcommand\thefigure{3}
\begin{figure}[t]
\begin{center}
\includegraphics[scale=0.6]{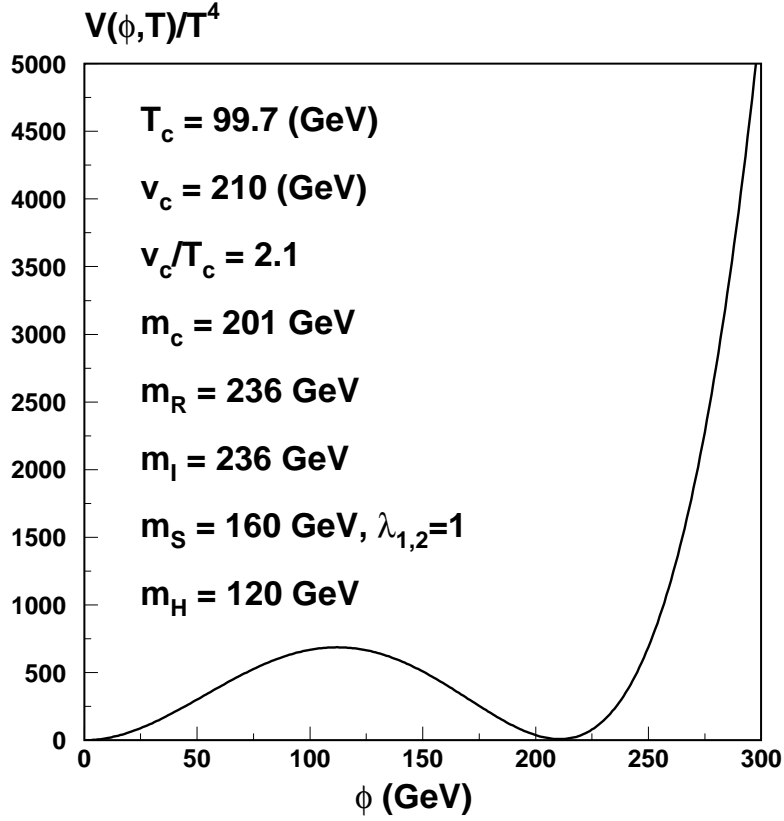}
\caption[plot]{The shape of the Higgs potential, for $\lambda_1 = \lambda_2 = 1$, $m_H = 120$ GeV,
and $m_S = 160$ GeV, when the critical temperature is $T_c = 99.7$ GeV.
The critical VEV is obtained as $v_c = 210$ GeV, yielding the strength of the first-order EWPT is about 2.1.
The masses of the scalar color octet are $m_{S_C} = 201$ GeV and  $m_{S_R} = m_{S_I} = 236$ GeV.
}
\end{center}
\end{figure}

\renewcommand\thefigure{4}
\begin{figure}[t]
\begin{center}
\includegraphics[scale=0.6]{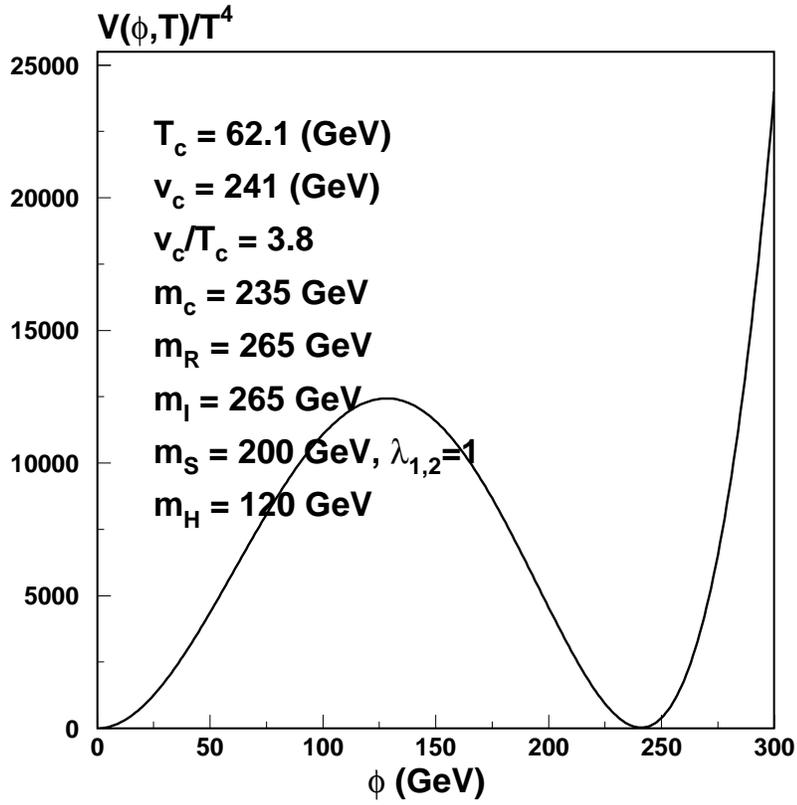}
\caption[plot]{The same as Fig. 3, except for $m_S = 200$ GeV, when $T_c = 62.1$ GeV.
The critical VEV is obtained as $v_c = 241$ GeV, and the strength of the first-order EWPT is about 3.8.
The masses of the scalar color octet are $m_{S_C} = 235$ GeV and $m_{S_R} = m_{S_I} = 265$ GeV.
}
\end{center}
\end{figure}

\renewcommand\thefigure{5}
\begin{figure}[t]
\begin{center}
\includegraphics[scale=0.6]{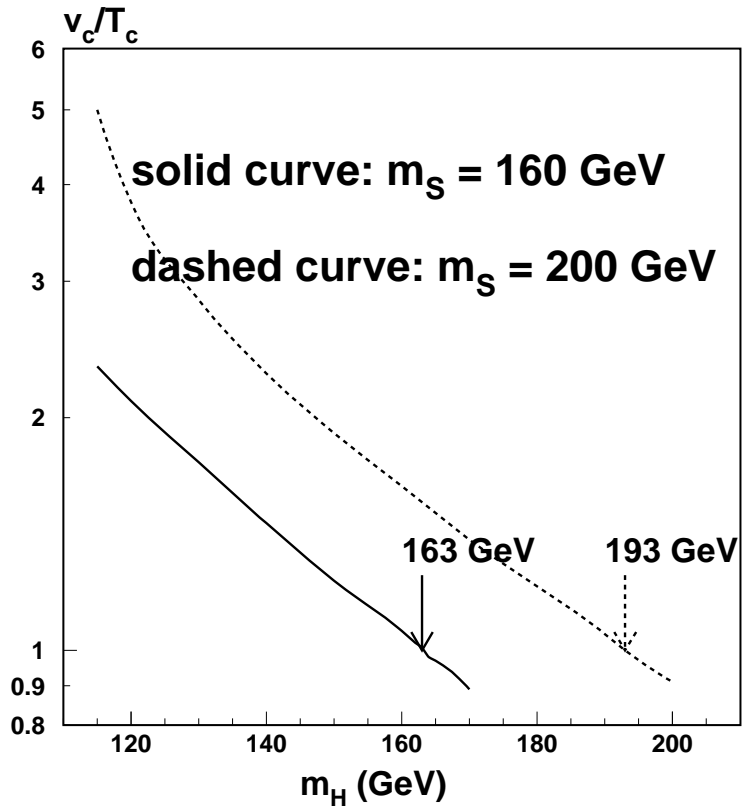}
\caption[plot]{The plots of the strength of the first-order EWPT as a function of $m_H$
for $m_S = 160$ GeV (solid curve) and for $m_S = 200$ GeV (dashed curve),
where $\lambda_1 = \lambda_2 = 1$.
Notice that for $m_S = 160$ GeV, the first-order EWPT is strong, i.e., $v_c/T_c > 1$,
for $m_H < 163$ GeV,
and for $m_S = 200$ GeV, it is strong for $m_H < 193$ GeV.
}
\end{center}
\end{figure}

\renewcommand\thefigure{6}
\begin{figure}[t]
\begin{center}
\includegraphics[scale=0.6]{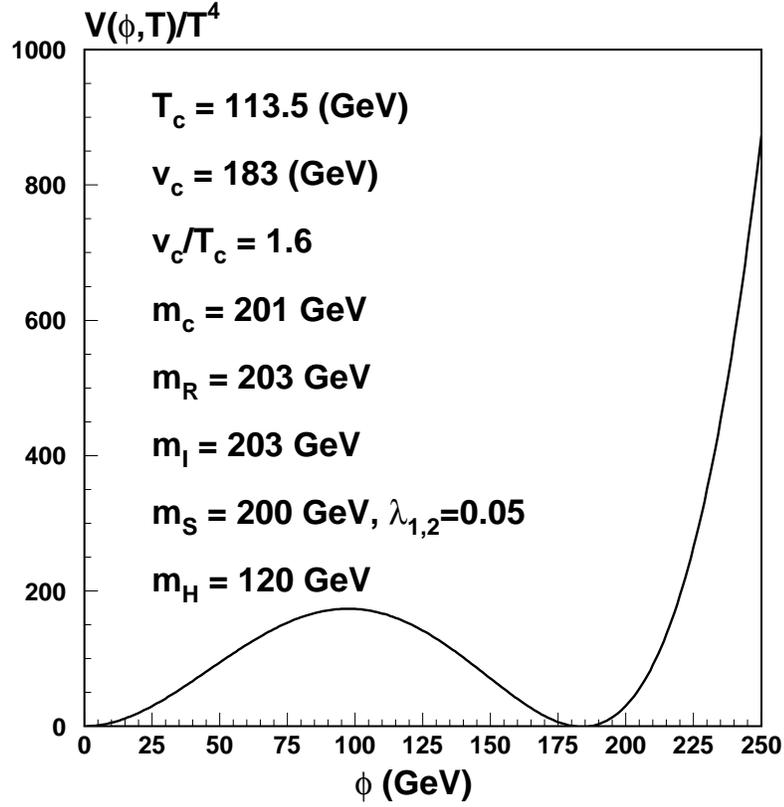}
\caption[plot]{The same as Fig. 3, except for $\lambda_1 = \lambda_2 = 0.05$ and $m_S = 200$ GeV,
when the critical temperature is $T_c = 113.5$ GeV.
The critical VEV is obtained as $v_c = 183$ GeV, yielding the strength of the first-order EWPT is about 1.6.
The masses of the scalar color octet are $m_{S_C} = 201$ GeV and $m_{S_R} = m_{S_I} = 203$ GeV.
}
\end{center}
\end{figure}

\renewcommand\thefigure{7}
\begin{figure}[t]
\begin{center}
\includegraphics[scale=0.6]{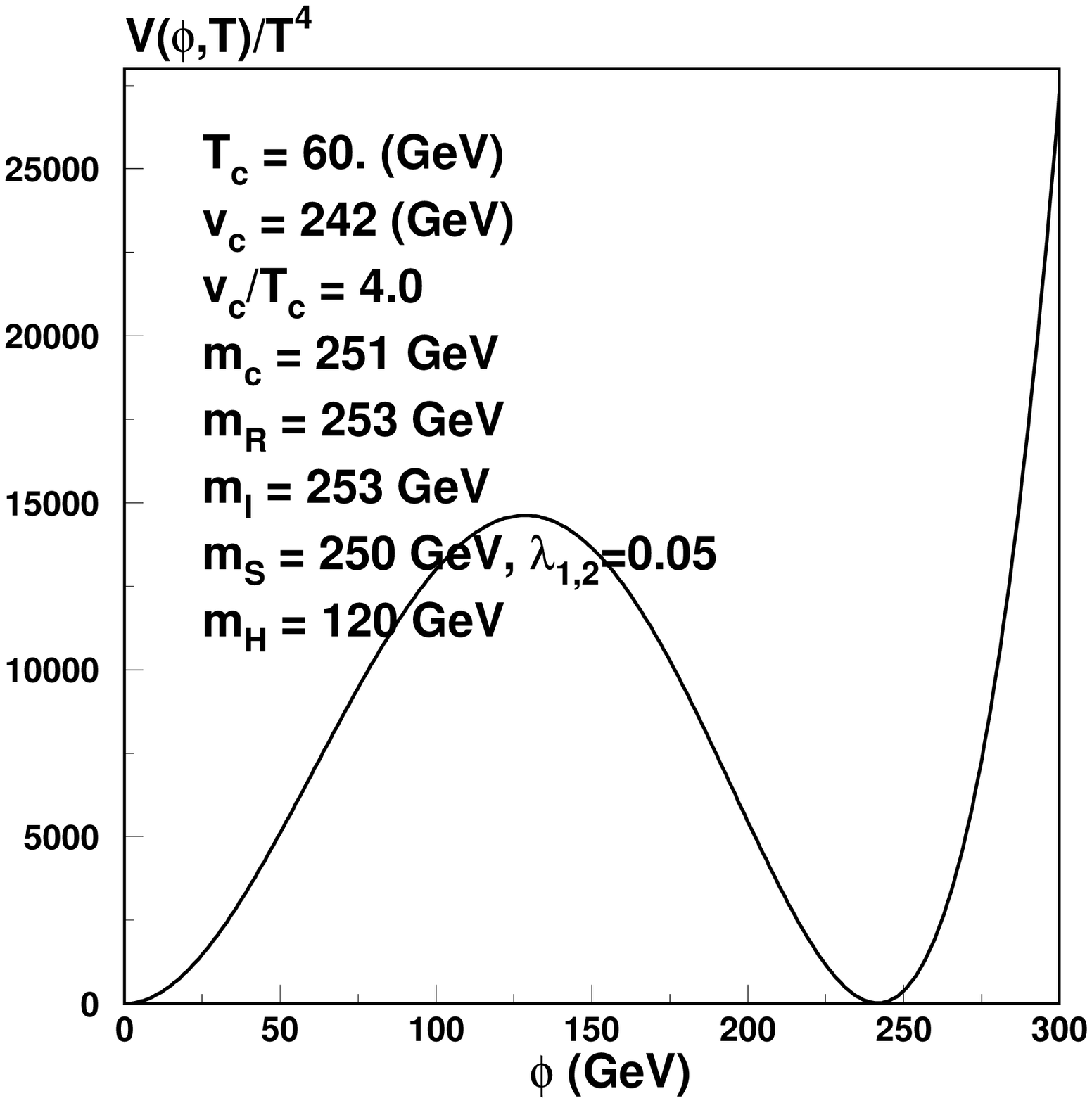}
\caption[plot]{The same as Fig. 6, except for $m_S = 250$ GeV, when $T_c = 60$ GeV.
The critical VEV is obtained as $v_c = 242$ GeV, yielding the strength of the first-order EWPT is about 4.0.
The masses of the scalar color octet are $m_{S_C} = 235$ GeV and $m_{S_R} = m_{S_I} = 265$ GeV.
}
\end{center}
\end{figure}

\renewcommand\thefigure{8}
\begin{figure}[t]
\begin{center}
\includegraphics[scale=0.6]{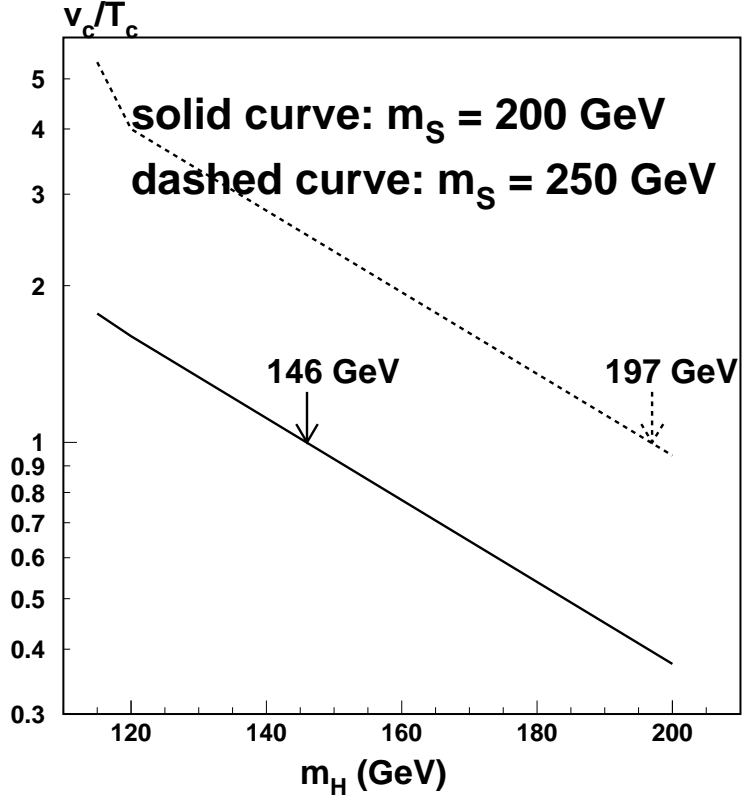}
\caption[plot]{The plots of the strength of the first-order EWPT as a function of $m_H$
for $m_S = 200$ GeV (solid curve) and for $m_S = 250$ GeV (dashed curve),
where $\lambda_1 = \lambda_2 = 0.05$.
Notice that for $m_S = 200$ GeV, the first-order EWPT is strong, i.e., $v_c/T_c > 1$,
for $m_H < 146$ GeV,
and for $m_S = 250$ GeV, it is strong for $m_H < 197$ GeV.
Fig. 8 may be compared with Fig. 5.
}
\end{center}
\end{figure}

\renewcommand\thefigure{9}
\begin{figure}[t]
\begin{center}
\includegraphics[scale=0.6]{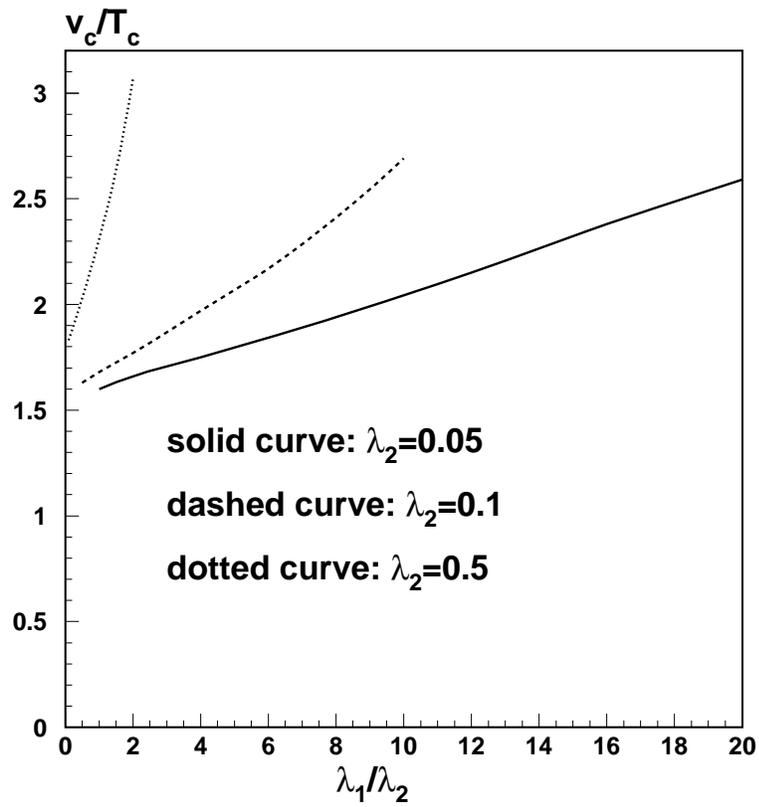}
\caption[plot]{The plots of the strength of the first-order EWPT as a function of $\lambda_1/\lambda_2$
for $\lambda_2 = 0.05$ (solid curve), for $\lambda_2 = 0.1$ (dashed curve), and
for $\lambda_2 = 0.5$ (dotted curve).
The values of the other free parameters are the same as in Fig. 6.
}
\end{center}
\end{figure}

\end{document}